\documentclass[preprint,aps,prd,showpacs,nofootinbib,superscriptaddress,floatfix,tightenlines]{revtex4-1}
\usepackage{amsmath}
\usepackage{graphicx}
\usepackage{subfigure}
\usepackage{epstopdf}
\usepackage{epsfig}
\usepackage{amssymb}
\usepackage{bm}
\usepackage{bbm}
\newcommand{\beq}{\begin{equation}}
\newcommand{\eeq}{\end{equation}}

\usepackage{color}

\begin{document}

\title{Dark Matter Production
\\
Associated With a Heavy Quarkonium at $\bm{B}$ Factories}

\author{Chaehyun Yu}
\email[]{chaehyun@gate.sinica.edu.tw}
\affiliation{Institute of Physics, Academia Sinica, Nangang, Taipei 11529, 
Taiwan}

\author{Tzu-Chiang Yuan}
\email[]{tcyuan@gate.sinica.edu.tw}
\affiliation{Institute of Physics, Academia Sinica, Nangang, Taipei 11529, 
Taiwan}

\date{\today}

\begin{abstract}
\noindent
We investigate light dark matter production associated with a heavy quarkonium
at $B$ factories in a model-independent way by adopting the effective field theory 
approach for the interaction of dark matter with standard model particles. 
We consider the effective operators for the dark matter-heavy quark interaction,
which are relevant to the production of dark matter associated with a heavy quarkonium.
We calculate the cross sections for dark matter production associated
with a $J/\psi$ or $\eta_c$ to compare with the standard model backgrounds.
We set bounds on the energy scale of new physics for various effective operators 
and also obtain the corresponding limits for the dark matter-nucleon scattering cross sections 
for light dark matter with mass of the order of a few GeV.

\end{abstract}

\maketitle

\section{Introduction}
\label{introduction}

Many astrophysical observations provide evidences for the existence 
of non-baryonic dark matter from the galactic scale to the cosmological 
scale~\cite{susydm,evidence,dm}. 
Its existence in our universe has been revealed only 
by its gravitational effects, while its mass and interactions to the standard
model (SM) particles are still unclear.
Among many scenarios that had been proposed to account for non-baryonic dark matter,
the weakly interacting massive particle (WIMP) scenario, 
where the particle with weak scale mass and interactions
could be thermally frozen out and its abundance would be the observed
dark matter relic density~\cite{susydm,evidence},
is one of the most attractive scenarios because it may be produced and detected
at the current or future
colliders. Actually many works in the literature have dealt with this 
issue~\cite{collider}, 
but they have mainly focused on the Large Hadron Colliders (LHC).

The dark matter search at colliders is based on the production of SM particle(s)
with missing energies. The typical signals are 
mono-jet~\cite{effective1,effective2,monophoton},
mono-photon~\cite{monophoton}, mono-$W$~\cite{monow} 
or mono-$Z$~\cite{monoz} with missing energies.
In principle, the particles which are produced together with the dark matter
particle(s) could be any SM fundamental particles or their composite states.
For example, the dark matter production associated 
with a heavy-quark-and-heavy-anti-quark ($Q\bar{Q}$) pair
may also play an important role in searching for dark matter.
It is known that the dark matter pair production associated with the $Q\bar{Q}$
pair could be more effective in the case that
the dark matter interaction with the SM particles depends on 
the mass of the relevant SM quark~\cite{monottbar}.

In this paper, we propose to use the process of
dark matter production associated with a heavy quarkonium 
to search for dark matter at colliders.
A heavy quarkonium is a meson which is a bound state of 
a $Q\bar{Q}$ pair. 
Naively speaking, the heavy-quark-pair-associated production of dark matter would be 
more effective for the dark matter detection than the light-quark-pair-associated
production if the effective coupling of dark matter to the quark pair
is proportional to the quark mass. 
In a similar way,  the dark matter production associated with a heavy quarkonium
would play a more essential role in the probe of such operators 
than dark matter production associated with a light meson.

The dark matter production associated with the SM particle(s) can be
investigated in any collider experiments.
In this work, we focus on the dark matter production
associated with a heavy quarkonium at $B$ factories, 
whose center-of-mass (CM) energy is $\sqrt{s}=10.58$ GeV. 
However, we note that this search will easily be
applied to the International Linear Collider (ILC) and Large Hadron Collider
(LHC)~\cite{future}.

There is no evidence for dark matter from the collider experiments so far.
However, there are some reports for dark matter candidate signals
from direct detection of dark matter, 
which is carried out in underground experiments,
and also from indirect detection of dark matter from astrophysical observations,
in particular, in the dark matter mass region of $\sim O(10)$ GeV.
On the other hand, the dark matter mass region below a few GeV 
has not been well investigated, especially, in the direct detection 
experiments of dark matter because the very light dark matter cannot 
hit a nucleon with significant recoil. 
One of the strongest bounds on the dark matter signals 
in the indirect detection of dark matter comes from the $\gamma$-ray
observation of  dwarf spheroidal galaxies 
by the Fermi-LAT satellite~\cite{dwarf}. 
However, the constraints become weakened below the dark matter mass region
less than about 4 GeV because of the uncertainty of hadronization 
of final state particles in the dark matter annihilation
and the limit of the photon energy threshold of the detector in the Fermi Gamma-ray Space
Telescope~\cite{Hdecay}.

Since such light dark matter mass region could be investigated
at colliders~\cite{light}, the collider detection of dark matter would be complementary 
to the direct detection of dark matter and the indirect detection 
of dark matter.

At $B$ factories, dark matter cannot only be produced directly,
but also produced from the decays of a heavy quarkonium.
For example, the invisible decay of $\Upsilon(1S)$ can give constraints
on the properties of light dark matter through the decay of
$\Upsilon(3S) \to \pi^+ \pi^- \Upsilon(1S), \gamma\Upsilon(1S)$ 
followed by $\Upsilon(1S) \to $ 
nothing at $B$ factories~\cite{Hdecay}.  
Also, the charm factories can contribute 
to search for dark matter by investigating the invisible decay of a charmonium,
$J/\psi$~\cite{Hdecay}.
It is found that the $\Upsilon(1S)$ and $J/\psi$ decays at $B$ and charm 
factories are more suitable to probe
light dark matter interactions than the mono-jet search at high energy hadron
colliders when the mass of mediator is not large~\cite{Hdecay}.
The dark matter production associated with a heavy quarkonium at $B$ factories
would also be complementary to the invisible decay of a heavy quarkonium
as well as the mono-jet searches at high energy colliders.
 
This paper is organized as follows. 
In Sec.~\ref{model}, we describe the relevant effective operators 
for the dark matter
interactions with the SM quarks. 
As a candidate for dark matter, we take into account two cases: Dirac
fermionic dark matter and real scalar dark matter.
We adopt the effective field theory (EFT) approach for the interaction of dark matter
with the SM quarks. 
In Sec.~\ref{calc}, we calculate the cross sections for
the dark matter production at $B$ factories
as well as for the SM backgrounds. We also interpret the constraints
on the properties of the dark matter interaction
to the upper limit on the dark matter-nucleon scattering cross section.
Finally, we summarize our results in Sec.~\ref{summary}.


\section{Effective operators}
\label{model}

For a model-independent search for dark matter, we assume that
$\chi$, which stands for either a Dirac fermion or real scalar, is
the only component of dark matter and singlet under the SM gauge group. 
The extension to the Majorana fermion
or complex scalar dark matter would be straightforward and not so much
different from the Dirac fermion or real scalar studied here.
We adopt the effective field
theory approach, where the mediator connecting the dark sector to the SM 
particles is heavy enough to be integrated 
out~\cite{effective1,effective2,effective3}.
After symmetry breaking, the dark matter interactions to the SM fields
can be described by higher-dimensional operators.
Since we are interested in the dark matter production associated 
with a heavy quarkonium, the relevant effective operators must contain
the heavy quark fields.

The dimension-6 operators in the Dirac fermion case which we will consider 
in this paper are~\cite{effective2}
\begin{eqnarray}
O_{1(2)}^D = (i)\frac{m_Q}{\Lambda^3} \bar{\chi} (\gamma^5)\chi \bar{Q} Q,
&~~&
O_{3(4)}^D = (-i)\frac{i m_Q}{\Lambda^3} \bar{\chi}(\gamma^5) \chi \bar{Q}\gamma^5  Q,
\nonumber \\
O_{5(6)}^D = \frac{1}{\Lambda^2} \bar{\chi}\gamma_\mu (\gamma^5) \chi 
\bar{Q}\gamma^\mu  Q,
&~~&
O_{7(8)}^D = \frac{1}{\Lambda^2} \bar{\chi}\gamma_\mu (\gamma^5) \chi 
\bar{Q}\gamma^\mu \gamma^5 Q,
\nonumber \\
O_{9}^D = \frac{1}{\Lambda^2} \bar{\chi}\sigma_{\mu\nu} \chi 
\bar{Q}\sigma^{\mu\nu}  Q,
&~~&
O_{10}^D = \frac{i}{\Lambda^2} \bar{\chi}\sigma_{\mu\nu}\gamma^5 \chi 
\bar{Q}\sigma^{\mu\nu}  Q,
\label{effopfermion}%
\end{eqnarray}
where $m_Q$ is the heavy-quark mass and
$\Lambda$ represents the energy scale of new physics.
Effectively $\Lambda \sim M /\sqrt{g_1 g_2}$, where $M$ is the mass
of the mediator and $g_i$ are the coupling of the mediator to
the dark sector and SM sector, respectively.
For the (pseudo)scalar interaction operators $O_{1,2,3,4}^D$,
an additional helicity suppression factor is taken into account.

In the real scalar case, the relevant dimension-6 operators 
are given by~\cite{effective2}
\begin{eqnarray}
O_{1}^R = \frac{m_Q}{2\Lambda^2} \chi^2 \bar{Q}Q,
&~~&
O_{2}^R = i\frac{m_Q}{2\Lambda^2} \chi^2 \bar{Q}\gamma^5 Q.
\label{effopscalar}%
\end{eqnarray}

There are more dimension-6 operators which would be necessary
for complete study of the dark matter production. 
For example, one can consider the operators which include 
the gluonic field strengths ($G_{\mu\nu}G^{\mu\nu}$).
They may contribute to the production of dark matter particles associated
with a heavy quarkonium at the one-loop level so that the dark matter
production process would be suppressed. 
For the probe of the operators which include the electromagnetic field strengths
($F_{\mu\nu}F^{\mu\nu}$) or a lepton pair, 
the dark matter production associated with a photon would be more efficient
at the lepton colliders like $B$ factories or ILC.

It is well known that the EFT approach is valid 
when the particle mediating the interaction is heavier than
the typical energy scale of the process, which is the maximum momentum transfer
in the process.
At high-energy hadron collider where the energy scale of the fundamental process 
is not fixed, the EFT approach may break down unless the mediator is 
not too heavy compared to the CM energy of the collider. 
Then, the EFT approach may over- or under-estimate the result in a UV-complete theory~\cite{break}.
It is worthwhile to note that $B$ factories are relatively free from
this issue because the CM energy is fixed as well as not so high.

Some operators in Eqs.~(\ref{effopfermion}) and (\ref{effopscalar})
do not respect the SM gauge group symmetry~\cite{violate}.
In particular, the scalar operators $O^D_{1,2,3,4}$ require
a $s$- or $t$ channel scalar exchange for UV completion.
In Ref.~\cite{monojetsf}, it is shown that in the $s$-channel scalar-exchange
model, the EFT approach prediction is completely different from that
of the UV complete model.
Therefore, one should be more cautious about the interpretation of the bounds on such operators
obtained in the EFT approach.


\section{Calculation}
\label{calc}

\subsection{Standard Model Backgrounds}

The production of a heavy quarkonium is described by non-relativistic QCD
(NRQCD), which is an effective field theory of QCD~\cite{nrqcd}.
As usual, the factorization between short-distance and long-distance physics is assumed. 
While the factorization between high-energy and low-energy physics 
for the inclusive heavy-quarkonium production has not been proved, the factorization for its exclusive production at $B$ factories was given in~\cite{factorization}.

The schematic expression for the production cross section is 
given by $d\sigma \sim \sum c_O \langle O \rangle_H$, where
$\langle O \rangle_H$ is the long-distance matrix element (LDME),
which is the probability of the evolution of a heavy-quark pair into a heavy 
quarkonium $H$,
and $c_O$ is the corresponding short-distance coefficient, which is 
responsible for the production of a heavy-quark and anti-quark pair.
The coefficient $c_O$ is obtained by integrating the squared amplitude over 
the phase space after averaging the spin states of initial particles
and summing the spin states of final particles.
The amplitude for the production of $H$ can be obtained by
projecting the production amplitude of a $Q\bar{Q}$ pair
onto $H$ via the spin and color projection operators of $H$ (generically denoted by $\Pi_H$ here):
\begin{equation}
{\cal M} \sim \textrm{Tr}[{\cal A} \Pi_H],
\end{equation}
with ${\cal A}$ being a matrix that acts on spinors of 
the $Q\bar{Q}$ pair in the amplitude of the $Q\bar{Q}$ pair production~\cite{BL1,resum}. 

In NRQCD, a heavy-quark and anti-quark pair can be produced in a color-singlet
state or color-octet state. The color-octet state can evolve into 
a color-singlet state by emitting or absorbing soft gluons.
Typically, the color-octet LDME is suppressed by $v^3$ or higher
in the heavy-quark velocity $v$ compared to the color-singlet LDME.
Therefore, unless there exists a kinematical enhancement factor in the short-distance
coefficient for the color-octet amplitude,
the color-singlet amplitude dominates.
In the $B$ factories, the initial colliding beams are the electron
and positron, which are neutral under $SU(3)_C$ gauge symmetry. 
Therefore, the production of the heavy quarkonium in the color-singlet state 
can occur only when the other particles are color-neutral or 
in the color-singlet state.
However, the production of a heavy quarkonium in a color-octet state
must accompany with at least a color-octet particle, 
for instance, an additional gluon or a pair of quark 
and anti-quark in the color-octet state.
In the dark matter production or off-shell $Z$ boson production at $B$ 
factories, the color-singlet heavy quarkonium production is dominant.
The color-octet heavy quarkonium production has a wavefunction suppression
factor of $O(v^3)$ or higher.
Furthermore, the color-octet process has an additional suppression factor 
due to the additional color-octet gluon or a pair of quark and anti-quark.
Therefore, for the dark matter production associated with a heavy quarkonium
at $B$ factories, it is sufficient to take into account
only the color-singlet contribution.

The color-singlet LDME is determined by the electromagnetic decay of the heavy
quarkonium. Explicitly, for charm quark mass $m_c=1.5$ GeV  and to leading-order
in $v$, we take the color-singlet LDMEs
to be~\cite{LDME,LDME2}
\begin{eqnarray}
\langle O \rangle_{\eta_c}  = 0.474~\textrm{GeV}^3,
&~~&
\langle O \rangle_{J/\psi}  = 0.436~\textrm{GeV}^3.
\end{eqnarray}

\begin{figure}[!t]
\begin{center}
{\epsfig{figure=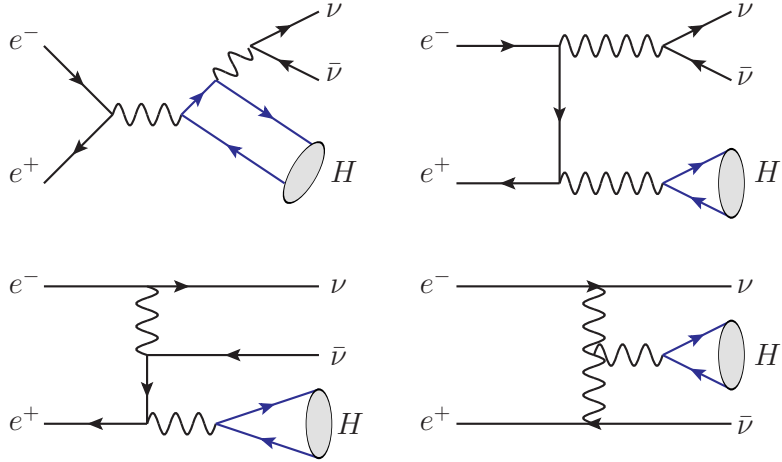,width=0.7\textwidth}}
\end{center}
\vspace{-0.5cm}
\caption{
The Feynman diagrams for $e^+ e^- \to H (J/\psi,\eta_c) \nu \bar{\nu}$ in the SM.
Other diagrams are obtained by reversing the flow of fermions.
}
\label{diagramsm}
\end{figure}

First we calculate the SM backgrounds.
In the SM, the Feynman diagrams contributing to the heavy quarkonium
production with missing energy have at least one weak-boson exchange,
which is connected to a pair of neutrinos.
There are also  diagrams with two weak-boson exchanges, but they
are suppressed by $O(s/M_Z^2)$ so that we ignore those diagrams.
Then, we left with seven diagrams in the SM, whose
representative diagrams are shown in Fig.~\ref{diagramsm}.
The irreducible background for the dark matter production
associated with a heavy quarkonium is the $e^+ e^- \to H Z^{\ast}$ process
followed by $Z^{\ast}\to \nu\bar{\nu}$, as depicted in the first two diagrams in Fig.~\ref{diagramsm}. 
There is also a $W$ fusion diagram, which is required by gauge invariance, 
as shown in the last diagram in Fig.~\ref{diagramsm}. 
Other diagrams not shown are obtained by reversing the flow of fermions.

In the monojet production at hadron colliders, except the irreducible 
SM background $pp\to Z+\textrm{jets}$ followed by $Z\to \nu\bar{\nu}$, 
there is another type of the SM background, for example, 
the process $pp\to W+\textrm{jets}, W\to l \nu$, where both $\nu$ and
$l$ are missed. We note that this kind of background can be ignored
at $B$ factories because the relevant process is
$e^+ e^- \to H W^{+\ast} W^{-\ast}$, which is highly suppressed. 

We note that, at the $e^+ e^-$ machines, there may be continuum background coming from $e^+ e^- \to q \bar{q} \nu\bar{\nu}$.
We calculate this continuum background by choosing
the invariant mass  $m_{q\bar{q}}$ of the $q\bar{q}$ pair in the region 
$m_{H} - 5 \Gamma_H \le m_{q\bar{q}} \le m_H + 5 \Gamma_H$, where
$m_H$ and $\Gamma_H$ are the mass and decay width
of $H$ respectively for $H=J/\psi$ or $\eta_c$. We find that the cross section for the continuum background is
less than $3.5\times 10^{-3}$ ab, which can be neglected at $B$ factories.
 There may also be a continuum background coming from process like $e^+ e^- \to \ell^+ \ell^- \nu\bar{\nu}$ ($\ell= e, \mu$)
 since $J/\psi$ may be detected experimentally via the decay $J/\psi \to \ell^+ \ell^-$.
Nevertheless the cross section from this continuum contribution is also negligible. 

We take the electromagnetic coupling to be $\alpha=1/132$ at the scale
$\mu=\sqrt{s}$ and the charm quark mass to be $m_c=1.5$ GeV.
Then, the cross sections for a heavy quarkonium production associated
with a pair of neutrinos in the SM are calculated as (1 ab = 1 attobarn = $10^{-42}$ cm$^2$)
\begin{eqnarray}
\sigma(e^+ e^- \to J/\psi \nu\bar{\nu}) &=& 0.81~\textrm{ab},
\\
\sigma(e^+ e^- \to \eta_c \nu\bar{\nu}) &=& 2\times 10^{-5}~\textrm{ab}.
\end{eqnarray}
The cross section for $\eta_c\nu\bar{\nu}$ production is much smaller than
that for the $J/\psi\nu\bar{\nu}$ production.
This is mainly because of the dominated contributions from photon-fragmentation diagrams, where
a virtual photon evolves into a $J/\psi$~\cite{resum,LDME2,photonfrag}.
All the diagrams in Fig.~\ref{diagramsm} contribute 
to the $J/\psi \nu\bar{\nu}$ production
while only the first diagram contributes to the $\eta_c \nu\bar{\nu}$
production. However, the first diagram does not correspond 
to the photon-fragmentation contribution and is suppressed by $O(m_c^2/s)$  
as compared to other photon-fragmentation diagrams.

The expected event number for the $J/\psi\nu\bar{\nu}$ production is about
$1$ $(40)$ with an integrated luminosity of $1$ $(50)$ ab$^{-1}$
without any cut. Since the expected event number for the $\eta_c\nu\bar{\nu}$
production is entirely negligible, even one event of the $\eta_c$ plus missing energy 
measured at $B$ factories would imply the existence of new physics.

\begin{figure}[!t]
\begin{center}
{\epsfig{figure=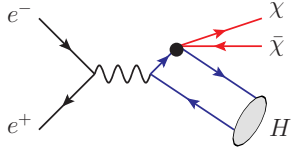,width=0.3\textwidth}}
\end{center}
\vspace{-0.5cm}
\caption{
The Feynman diagram for $e^+ e^- \to J/\psi + \chi +\bar{\chi}$.
Another diagram is obtained by reversing the flow of the charm quark.
The bulb represents an effective operator.
}
\label{diagram}
\end{figure}

\subsection{The Dirac Fermion Dark Matter Case}

In this section, we calculate the cross section for the Dirac fermion dark matter production
associated with a heavy quarkonium. 

There are two Feynman diagrams, one of
which is shown in Fig.~\ref{diagram}. The other diagram is obtained
by reversing the flow of the charm quark.
The bulb in the figure represents vertices of various effective operators.

The relevant operators to the heavy quarkonium production depend
on the quantum number of the heavy quarkonium,
where $J^{PC}=1^{--}$ for $J/\psi$ and $J^{PC}=0^{+-}$ for $\eta_c$.
For example, the operators $O_{1,2,3,4,7,8}^D$ contribute only to  
the $e^+ e^- \to J/\psi \chi \bar{\chi}$ process,
while the operators $O_{5,6,9,10}^D$ 
only to the $e^+ e^- \to \eta_c \chi\bar{\chi}$.
We note that only the $O_{5,6,9,10}^D$ operators
can contribute to the invisible decays of the heavy quarkonium states 
$\Upsilon(1S)$ and $J/\psi$~\cite{Hdecay}.
Thus, the dark matter production associated with a heavy 
quarkonium could provide constraints on a larger set of operators, which 
cannot be probed in the invisible decays of $\Upsilon(1S)$ and $J/\psi$.

\begin{figure}[!t]
\begin{center}
{\epsfig{figure=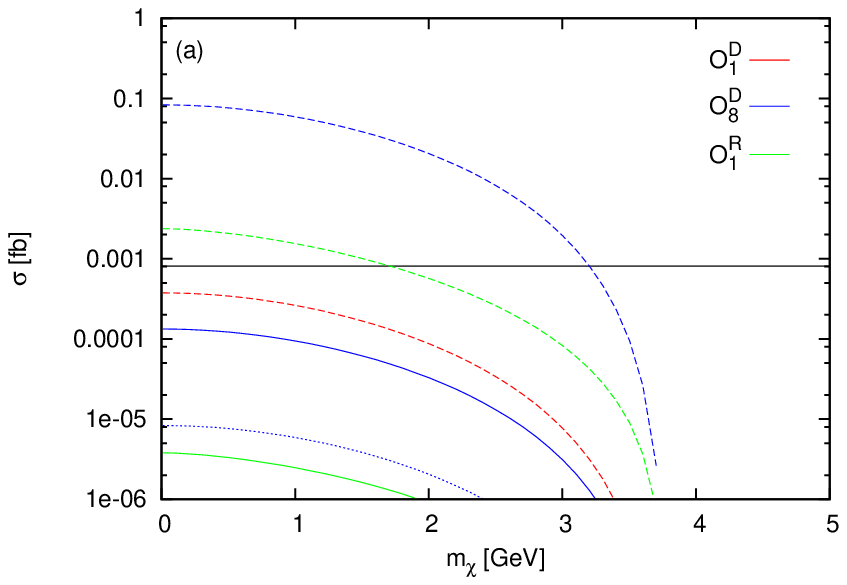,width=0.4\textwidth}}
{\epsfig{figure=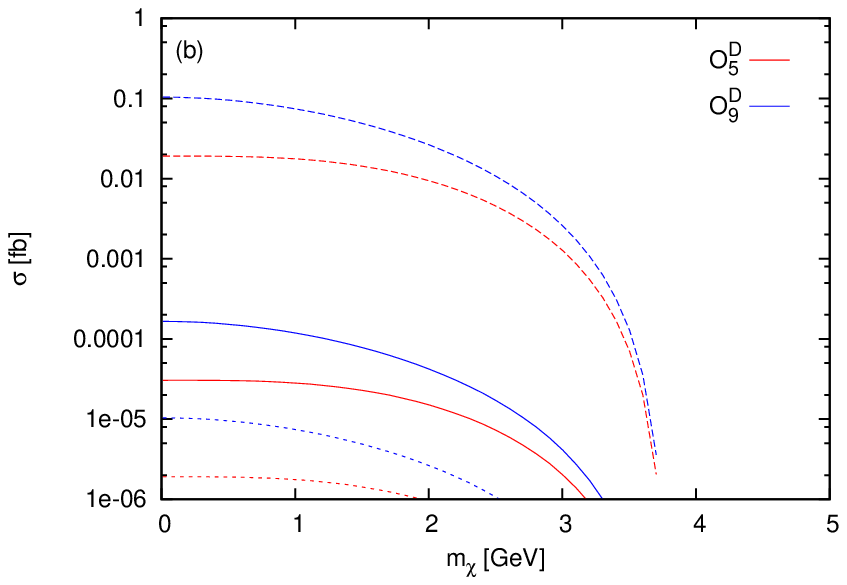,width=0.4\textwidth}}
\end{center}
\vspace{-0.5cm}
\caption{
The cross sections (a) for $e^+ e^- \to J/\psi  \chi \bar{\chi}$
and (b) for $e^+ e^- \to \eta_c \chi \bar{\chi}$
in unit of fb as a function of $m_\chi$ in unit of GeV.
The red and blue lines correspond to the $O_1^D (O_5^D)$ 
and $O_8^D (O_9^D)$
operators for the $J/\psi \chi \bar{\chi}$ ($\eta_c\chi\bar{\chi}$)
production, respectively.
The dashed, solid, and dotted lines
correspond to $\Lambda=20$, $100$, and $200$ GeV, respectively.
The green lines correspond to the $O_1^R$ operator for the real scalar dark
matter.
}
\label{figurexsec}
\end{figure}

Figure~\ref{figurexsec} shows the cross sections
(a) for $e^+ e^- \to J/\psi  \chi \bar{\chi}$ and 
(b) for $e^+ e^- \to \eta_c  \chi \bar{\chi}$  
as a function of the dark matter mass $m_\chi$ for $\Lambda=20$ GeV (dashed line),
$100$ GeV (solid line), and $200$ GeV (dotted line), respectively.
The effective operators for the red and blue lines 
are $O_1^D$ ($O_5^D$) and $O_8^D$ ($O_9^D$) for the $J/\psi$ ($\eta_c$)
production, respectively.
The horizontal line in Fig.~\ref{figurexsec}(a) is the SM background
in the framework of NRQCD 

In Fig.~\ref{figurexsec}(a), the cross sections 
for the $J/\psi \chi\bar{\chi}$ production can reach
about $0.08$ fb for $\Lambda= 20$ GeV and about $0.01$ ab for $\Lambda=100$ GeV
respectively in the case of the operator $O_8^D$. 
However, for the $O_1^D$ operator, the cross section
reaches about $0.4$ ab for $\Lambda=20$ GeV, but the production is negligible
for $\Lambda=100$ GeV due to a suppression factor $O(m_c/\Lambda)$.
The curves in the figure drop rapidly as the dark matter mass grows, indicating 
the kinematical limit of the dark matter mass of $(\sqrt{s}-m_{J/\psi})/2$.
For other operators $O_{2,3,4,7}^D$, similar features can be obtained.

In Fig.~\ref{figurexsec}(b), the cross section for the $\eta_c\chi\bar{\chi}$
production can reach about $0.02$ fb for $\Lambda=20$ GeV
and about $10$ ab for $\Lambda=100$ GeV in the case of the $O_5^D$ operator.
For the $O_9^D$ operator, the cross section reaches about
$0.1$ fb for $\Lambda=20$ GeV and about $0.17$ ab for $\Lambda=100$ GeV,
respectively. Effects from the kinematical limit of the dark matter mass of
$(\sqrt{s}-m_{\eta_c})/2$ are also evident.
For operators $O_{6,10}^D$, similar features are obtained.

\begin{figure}[!t]
\begin{center}
{\epsfig{figure=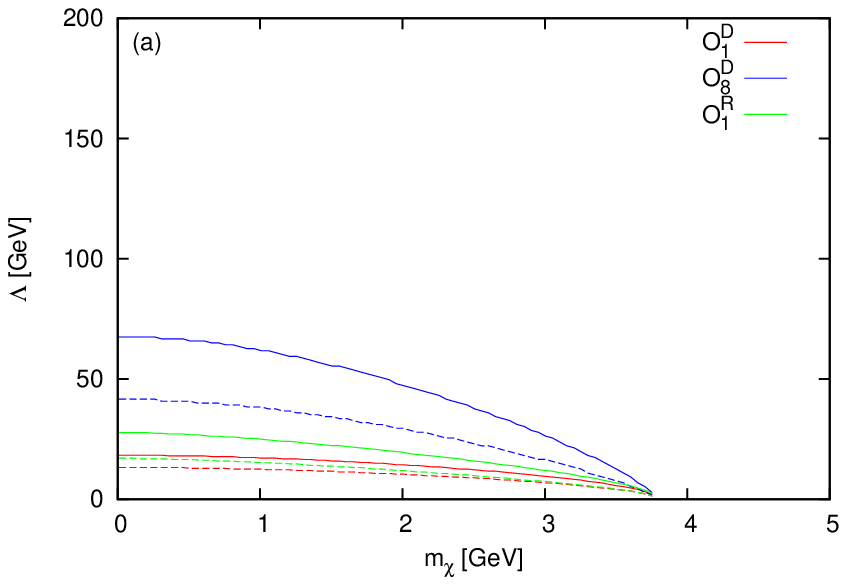,width=0.4\textwidth}}
{\epsfig{figure=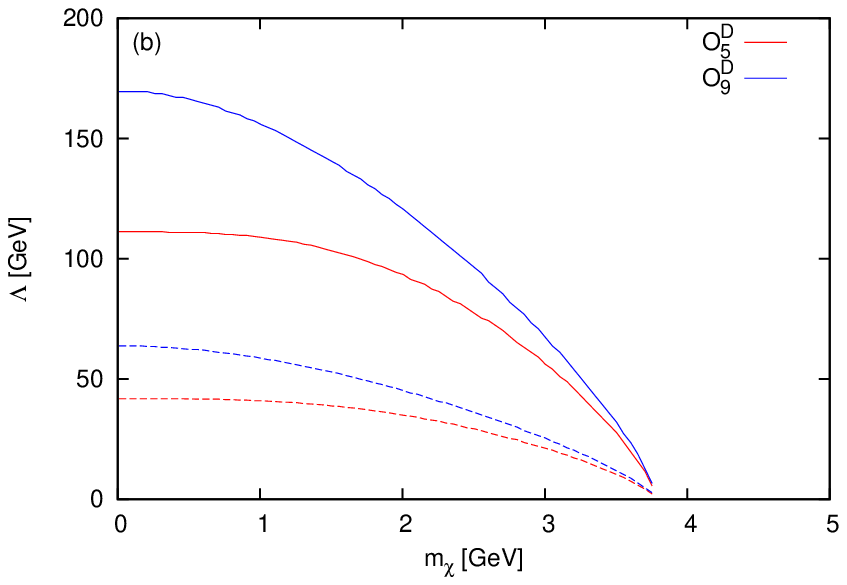,width=0.4\textwidth}}
\end{center}
\vspace{-0.5cm}
\caption{
Bounds on $\Lambda$ (a) for $e^+ e^- \to J/\psi  \chi \bar{\chi}$
and (b) for $e^+ e^- \to \eta_c \chi \bar{\chi}$
as a function of $m_\chi$, determined by the signal-to-background ratio $R=5$ and 1 respectively.
The red and blue lines correspond to the $O_1^D (O_5^D)$ and $O_8^D (O_9^D)$
operators in the $J/\psi \chi \bar{\chi}$ ($\eta_c\chi\bar{\chi}$)
production, respectively.
The green lines correspond to the $O_1^R$ operator in the $J/\psi \chi \chi$ 
production where $\chi$ is a real scalar.
The dashed and solid lines
correspond to the integrated luminosity of $1$ ${\rm ab}^{-1}$ and $50$ ${\rm ab}^{-1}$, respectively.
}
\label{figurelam}
\end{figure}

In Fig.~\ref{figurelam}, we plot bounds on the scale $\Lambda$
(a) for $e^+ e^- \to J/\psi \chi \bar{\chi}$
and (b) for $e^+ e^- \to \eta_c \chi \bar{\chi}$
in the cases of the integrated luminosity (${\cal L}$) of $1$ ${\rm ab}^{-1}$ (dashed line) 
and $50$ ${\rm ab}^{-1}$ (solid line), respectively.
In the $J/\psi \chi\bar{\chi}$ production, we set the bound on $\Lambda$
to be the value where the signal-to-background ratio, which is defined by
$R={\cal L} \sigma_\textrm{sig}/\sqrt{{\cal L}\sigma_\textrm{bg}}$, is equal
to 5.
In the $\eta_c\chi\bar{\chi}$ production, we set the bound
to be the value where the number of events is one because the SM background
is negligible.
In the $J/\psi\chi\bar{\chi}$ production, the bound can reach
about $42$ ($67$) GeV with ${\cal L}=1$ ($50$) ${\rm ab}^{-1}$ for the $O_8^D$ operator.
For the $O_1^D$ operator, the bound is much less due to the chirality suppression 
factor in the operator.
In the $\eta_c\chi\bar{\chi}$ production, the bound can reach
about $42$ ($111$) GeV for $O_5^D$ operator
and about $64$ ($169$) GeV for $O_9^D$ operator
with ${\cal L}=1$ ($50$) ${\rm ab}^{-1}$,  respectively.

A couple of comments are in order here.
First, as the dark matter mass approaches the kinematical limit, 
the bound on $\Lambda$ which can be achieved at $B$ factories becomes quite tiny
as clearly shown in Fig.~\ref{figurelam}.
We note that for the region below some value of $\Lambda$, our limits become untrustworthy. 
Recall that $\Lambda \sim M / \sqrt{g_1g_2}$. A very tiny $\Lambda$ can be achieved 
only when $g_1 g_2$ becomes large since
the mediator mass $M$ should be larger than $\sqrt{s}$ 
for validity of EFT approach. On the other hand, 
$g_{1,2}$ cannot be too large in order to perform perturbative calculation.
For $g_{1,2} \sim 1$, the EFT description breaks down  
in the region $\Lambda \lesssim \sqrt{s}$.
Second, we note that the number of events is not large. So the best thing 
one can hope for is to detect inclusive signals rather than exclusive ones. 
That is, the heavy quarkonium would be identified by the invariant mass distribution of 
all the final state particles in the decay products (except for missing energy) 
instead of using the exclusive mode like $J/\psi \to \l^+l^-$.

\subsection{The Scalar Dark Matter Case}

In this section, we consider the real scalar dark matter production associated
with a heavy quarkonium at $B$ factories. The relevant Feynman diagrams 
are the same as in the Dirac dark matter case with the replacement of
the dark matter fermion lines by the dark matter scalar lines.
The relevant effective operators are shown in Eq.~(\ref{effopscalar}).

In Fig.~\ref{figurexsec}(a), the cross section for the $J/\psi\chi\chi$ 
production in the $O_1^R$ operator case 
is depicted for $\Lambda=20$ GeV (green dashed line) and $\Lambda=100$ GeV
(green solid line). The cross section can reach about $2$ ab for $\Lambda=20$ GeV,
but for $\Lambda=100$ GeV the cross section is below $O(10^{-2})$ ${\rm ab}^{-1}$.
Similar feature is observed for the $O_2^R$ operator.
We note that the cross section for the $\eta_c \chi\chi$ production
vanishes for real dark matter.

The bound on the scale $\Lambda$ for $e^+ e^- \to J/\psi \chi \chi$
is shown in Fig.~\ref{figurelam}(a) for the integrated luminosity
of 1 ${\rm ab}^{-1}$ (green dashed line) and 50 ${\rm ab}^{-1}$ (green solid line), respectively.
Again the bound is set to be the signal-to-background ratio $R=5$.
The bound on $\Lambda$ can reach about $17$ $(28)$ GeV 
for ${\cal L}=1$ $(50)$ ${\rm ab}^{-1}$ respectively
for the operator $O^R_1$.

\begin{figure}[!t]
\begin{center}
{\epsfig{figure=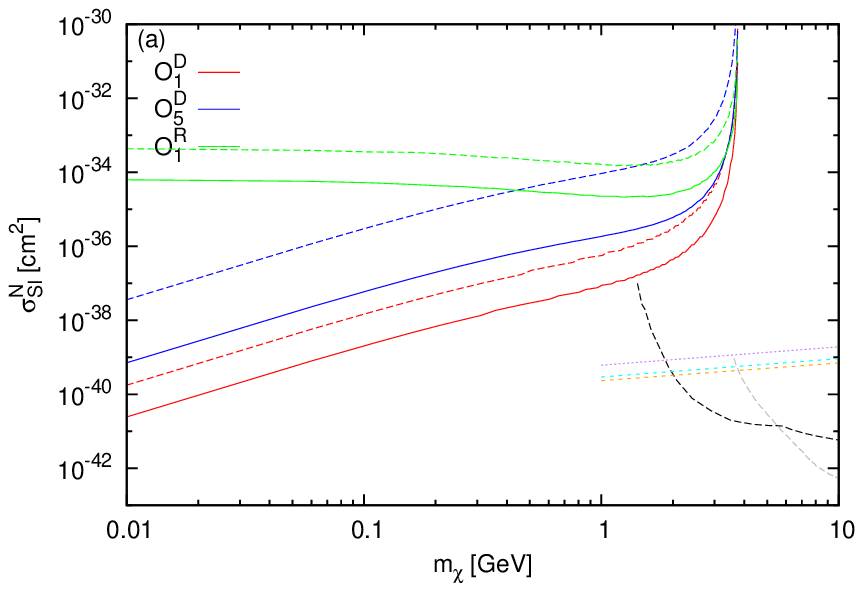,width=0.4\textwidth}}
{\epsfig{figure=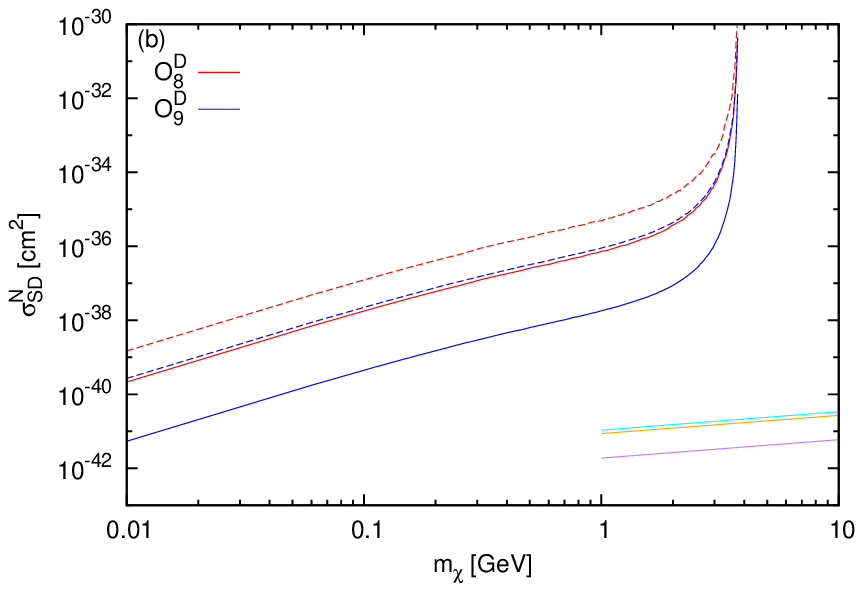,width=0.4\textwidth}}
\end{center}
\vspace{-0.5cm}
\caption{
Experimental reach of (a) the spin-independent cross section
and (b) the spin-dependent cross section 
for the dark matter-nucleon scattering in units of cm$^2$
as a function of $m_\chi$ in unit of GeV.
The red (blue) lines correspond to the $O_1^D (O_5^D)$ and $O_8^D (O_9^D)$
operators in the $J/\psi \chi \bar{\chi}$ ($\eta_c\chi\bar{\chi}$)
production, respectively.
The green lines in (a) correspond to the $O_1^R$ operator 
in the $J/\psi \chi \bar{\chi}$ production.
The dashed and solid lines
correspond to the integrated luminosity of $1$ ${\rm ab}^{-1}$ and $50$ ${\rm ab}^{-1}$, respectively.
The black and gray lines on the lower right corners are 90\% C.L. exclusion limits
of second CDMSlite run~\cite{cdmslite2} and superCDMS~\cite{supercdms}, respectively.
The bounds from the monojet search at the LHC are shown in 
the orange dotted (solid) lines for $O^D_5$ ($O^D_8$) from CMS,
the cyan dotted (solid) lines for $O^D_5$ ($O^D_8$) from ATLAS,
and the purple dotted (solid) lines for $O^D_1$ ($O^D_9$) from ATLAS, 
respectively~\cite{cmsmonojet,atlasmonojet} from 1 to 10 GeV.
}
\label{figuredirect}
\end{figure}

\subsection{Conversion to the Dark Matter-Nucleon Cross Section}

The bound on $\Lambda$ deduced above may be converted into the bound for
the dark matter-nucleon scattering cross section for the direct detection of dark matter.
We note that, however, this conversion holds with the assumption
that the effective operators in Eqs.~(\ref{effopfermion}) and (\ref{effopscalar})
are either flavor blind or proportional to the quark masses. 
In case that dark matter has charm-philic property, the bound from 
the dark matter production associated with a heavy quarkonium cannot be
converted into the cross section for the dark matter-nucleon scattering.

For the effective operators, the dark matter-nucleon scattering 
cross sections are shown in Refs.~\cite{belanger,Haisch:2012kf}:
\begin{eqnarray}
\label{dmnucleona}
\sigma_\textrm{SI}{(O^D_1)} &=& 
\frac{\mu_\chi^2 m_n^2}{\pi\Lambda^6}f_n^2,
\\
\label{dmnucleonb}
\sigma_\textrm{SI}{(O^D_5)} &=& 
\frac{\mu_\chi^2}{\pi \Lambda^4} \left(\sum_{q}f_{V_q}^n\right)^2,
\\
\label{dmnucleonc}
\sigma_\textrm{SI}{(O^R_1)} &=& 
\frac{4 \mu_\chi^2 m_n^2}{\pi \Lambda^4 m_\chi^2} f_n^2,
\\
\label{dmnucleond}
\sigma_\textrm{SD}{(O^D_{8,9})} &=& 
\frac{3 \mu_\chi^2}{\pi \Lambda^4} \left(\sum_{q} \Delta_q^n\right)^2,
\end{eqnarray}
where $\mu_\chi$ is the reduced mass of the dark matter and nucleon
and $m_n$ is the mass of the nucleon.
$\sigma_\textrm{SI,SD}$ stands for the spin-independent (SI)
and spin-dependent (SD) cross section for the dark matter-nucleon scattering.
The scalar form factor of the nucleon 
\begin{equation}
f_n = \sum_{q=u,d,s} f_q^n + \frac{2}{27} \sum_{Q=c,b,t} f_Q^n,
\end{equation}
where $f_Q^n=1-f_u^n-f_d^n-f_s^n$. Here, we use
$f_d^p = 0.017$, $f_u^p = 0.023$, and $f_s^p=0.053$, 
for the scalar form factors,
$f_{V_u}^p=2$, $f_{V_d}^p=1$, for the vector form factors, and
$\Delta_{u}^p=0.85$, $\Delta_{d}^p=-0.42$, and $\Delta_{s}^p=-0.08$
for the axial-vector form factors, respectively~\cite{formfactor,formfactor2}.
We note that if the dark matter particles interact only with the charm quarks,
the dark matter-nucleon scattering cross sections are negligible or significantly reduced.

In Fig.~\ref{figuredirect}, we show (a) the spin-independent cross section
and (b) the spin-dependent cross section versus the dark matter mass
with the bounds of $\Lambda$ extracted for each dark matter mass 
from Fig.~\ref{figurelam} for the corresponding operators and luminosities 
and applied to the appropriate formulas given in 
Eqs.~(\ref{dmnucleona}) to  (\ref{dmnucleond}).
The black and gray lines at the lower right corners are 90\% C.L. exclusion limits
of CMDSlite~\cite{cdmslite2} and superCDMS~\cite{supercdms}, respectively.
Also at the lower right corners of these two plots in Fig.~\ref{figuredirect}, 
the orange dotted (solid) lines are 90\% C.L. exclusion limits
of CMS for $O^D_5$ ($O^D_8$)  from 1 to 10 GeV~\cite{cmsmonojet},
while the cyan dotted (solid) lines are for $O^D_5$ ($O^D_8$) from ATLAS
and the purple dotted (solid) lines are for $O^D_1$ ($O^D_9$) from ATLAS, 
respectively~\cite{atlasmonojet}. Since the bounds from CMS and ATLAS 
are obtained for the effective operators (\ref{effopfermion}),
they strongly depend on the UV completion of the operators
if other particles in the UV completion model are not heavy enough to be integrated
out at the LHC energy scale.

The red and blue lines correspond to the $O_1^D (O_5^D)$
and $O_8^D (O_9^D)$ operators in the $J/\psi \chi\bar{\chi}$ 
($\eta_c \chi\bar{\chi}$) production, respectively. 
The green line corresponds to the $O_1^R$ operator in the $J/\psi \chi\chi$
production.
The dashed and solid lines represent the cross section corresponding
to ${\cal L}=1$ ${\rm ab}^{-1}$ and $50$ ${\rm ab}^{-1}$ at $B$ factories, respectively.
The bound on the dark matter-nucleon scattering cross section could reach about 
$10^{-41}$ cm$^2$ in the spin-independent case and 
$10^{-42}$ cm$^2$ in the spin-dependent case for the very light $m_\chi$.
For $m_\chi \sim 1 \, \textrm{GeV}$ and ${\cal L}=50$ ${\rm ab}^{-1}$ at $B$ factories, 
the bound could be about $\sim 10^{-38}$ cm$^2$ in the spin-independent case and
about $\sim 10^{-39}$ cm$^2$ in the spin-dependent case.
The bounds obtained from the mono-jet search at the LHC might be stronger
than those from $B$ factories. However, the bounds from the LHC are restricted
to the large mass region of the mediator, but those from $B$ factories
would be applied to broader range of the mediator mass 
($\gtrsim \sqrt{s}$).

As shown in Fig.~\ref{figuredirect}, the LHC bounds are more efficient
than the bounds obtained from the dark matter production associated
with a heavy quarkonium at $B$ factories by a few orders of magnitude
for $m_\chi > 1$ GeV. However, this search is meaningful to provide 
complementary search in the lepton colliders for the universal couplings
of dark matter.
If the dark matter particle dominantly couples to the charm quark,
the LHC bound would become weaker for $O_{5,8,9}^D$
since the current bounds arise from the light quark interactions
in the $q\bar{q}$ or $qg$ ($q=u,d$) collisions.
Furthermore, the bounds from the direct detection could be negligible
since they do not give any constraints on such operators at leading order.
For the scalar operators $O_1^{D,R}$, the LHC bounds are still powerful,
but the direct detection bounds become weaker by $O(0.01)$.

We note that one should be careful for the comparison of the results of dark matter search 
at colliders with those from direct detection.
As shown in Ref.~\cite{monojetsf}, the EFT approach may be 
inadequate for the hadron collider dark matter search because the typical energy
scale in the process is not fixed.
In addition, some of the effective operators in Eqs.~(\ref{effopfermion})
and (\ref{effopscalar}) do not respect the SM gauge symmetry,
in particular, $SU(2)_L$ symmetry.
If one imposes the SM gauge symmetry for the effective operators, 
the UV complete model, which is responsible for generating these operators,
may not approach the EFT even in the limit of infinite mass of the mediator.
This is true for the scalar operators in Eq.~(\ref{effopfermion})~\cite{monojetsf}.
The contact interaction for the scalar$\times$scalar operator cannot be realized
at the energy scale of the LHC because the SM Higgs boson which is needed to impose
the SM $SU(2)_L$ gauge symmetry is still a propagating mode.
In this case, the comparison in Fig.~\ref{figuredirect} should be
re-interpreted with the analysis in the full theory.
However, at $B$ factories, one can obtain the contact interaction 
by integrating out both the heavy degree of freedom of the mediator and SM Higgs boson.
The effective scale $\Lambda$ will depend on the masses of the mediator as well as 
the SM Higgs boson. The effective operators thus obtained would respect just
the gauge symmetries of QED and QCD, which is necessary.


\section{Summary}
\label{summary}

In this paper, we investigate light dark matter production associated with 
a heavy quarkonium at $B$ factories. For the interaction of dark matter
to the SM sector, the effective field theory approach is adopted, but
it would be straightforward to extend this study to a realistic model
with UV completion.
So far the dark matter search at colliders has been focused on hadron colliders.
In this paper, we showed that the $B$ factories may play a role in searching
for dark matter. 
Especially, for the light mediator mass, the $B$ factories would be more
effective. 
We took into account the $J/\psi \chi\bar{\chi}$ and $\eta_c\chi\bar{\chi}$
production, but this would easily be extended to the charmonium in 
the higher resonances like $\psi(2S)$, $\eta_c(2S)$, and $\chi_{cJ}$,
as well as the bottomonium production. Combining all the results, one may
obtain stronger bounds on the effective energy scale $\Lambda$,
which together with the dark matter mass determines the size of
the dark matter-nucleon scattering cross section for direct detection.

The EFT approach holds only when the mediator mass is larger than
the typical energy scale of the relevant processes.
Therefore the dark matter search via the EFT approach
would be reliable only for $M \gtrsim \sqrt{s}$.
For $M \lesssim \sqrt{s}$, one must consider a UV complete model
or another EFT model including the particles which cannot be integrated out
at the energy scale. 
Another important point for the EFT approach is to impose
the full SM gauge symmetry~\cite{monojetsf}. 
For example, the $SU(2)_L$ symmetry may require the SM Higgs boson
exchange between the dark sector and SM sector for the scalar$\times$scalar
interaction, which cannot be integrated out at the energy of the LHC.
Searches for dark matter at the $B$ factories are
relatively free from such complications. 

The search for light dark matter via its production associated with a heavy quarkonium
at $B$ factories would provide both the alternative and complementary ways
to the hadron colliders and the heavy quarkonium invisible decays.

\acknowledgments
This work was supported in part by the Ministry of Science and Technology (MoST) of Taiwan 
under grant number 101-2112-M-001-005-MY3.

\end{document}